%% file: AHASD.tex
\documentclass[sigconf, nonacm]{acmart}
\usepackage{subcaption}
\usepackage{multirow}
\makeatletter
\def\blfootnote{\gdef\@thefnmark{}\@footnotetext}
\makeatother
\renewcommand{\footnoterule}{\kern-3pt\hrule width\columnwidth\kern2.6pt}

\AtBeginDocument{%
  }

\begin{document}
\pagestyle{empty}

\title{AHASD: Asynchronous Heterogeneous Architecture for LLM Adaptive Drafting Speculative Decoding on Mobile Devices}

\author{Zirui Ma$^{1,2}$, Zhihua Fan$^{1}$*, Wenxing Li$^{1}$, Haibin Wu$^{1,2}$, Fulin Zhang$^{1,2}$, Wenming Li$^{1,2}$*, Xiaochun Ye$^{1,2}$}
\affiliation{%
  \institution{$^{1}$ State Key Lab of Processors, Institute of Computing Technology, Chinese Academy of Sciences}
  \city{}
  \country{}}
\affiliation{%
  \institution{$^{2}$ University of Chinese Academy of Sciences}
  \city{}
  \country{}}

\begin{abstract}

Speculative decoding enhances the inference efficiency of large language models (LLMs) by generating drafts using a small draft language model (DLM) and verifying them in batches with a large target language model (TLM). However, adaptive drafting inference on a mobile single-NPU-PIM system faces idle overhead in traditional operator-level synchronous execution and wasted computation in asynchronous execution due to fluctuations in draft length. This paper introduces AHASD, a task-level asynchronous mobile NPU-PIM heterogeneous architecture for speculative decoding. Notably, AHASD achieves parallel drafting on the PIM and verification on a single NPU through task-level DLM-TLM decoupling and specifically, it incorporates Entropy-History-Aware Drafting Control and Time-Aware Pre-Verification Control to dynamically manage adaptive drafting algorithm execution and pre-verification timing, suppressing invalid drafting based on low-confidence drafts. Additionally, AHASD integrates Attention Algorithm Units and Gated Task Scheduling Units within LPDDR5-PIM to enable attention link localization and sub-microsecond task switching on the PIM side. Experimental results for different LLMs and adaptive drafting algorithms show that AHASD achieves up to 4.2$\times$ in throughput and 5.6$\times$ in energy efficiency improvements over a GPU-only baseline, and 1.5$\times$ in throughput and 1.24$\times$ in energy efficiency gains over the state-of-the-art GPU+PIM baseline, with hardware overhead below 3\% of the DRAM area.

\end{abstract}





\renewcommand{\shortauthors}{Zirui Ma et al.}
\maketitle
\blfootnote{\textbf{Acknowledgement:} This work was supported by National Key R\&D Program of China (Grant No.2023YFB4503500), CAS Project for Young Scientists in Basic Research under Grant YSBR-029, National Natural Science Foundation of China (Grant No.62502498), Beijing Natural Science Foundation (Grant No.L234078).\\
*Zhihua Fan and Wenming Li are corresponding authors.}

\input{sections/section1.tex}
\input{sections/section2.tex}
\input{sections/section3.tex}
\input{sections/section4.tex}

\input{sections/section5.tex}
\input{sections/section6.tex}

\bibliographystyle{ACM-Reference-Format}
\bibliography{sample-base}

\appendix

\end{document}

%% file: sections/section1.tex
\section{Introduction}

Speculative decoding~\cite{decoding, draftVerify, fastInference, specInfer} offers a novel system and algorithm co-design approach to enhance the inference efficiency of large language models (LLMs) without significantly sacrificing generation quality. The core idea involves a small-scale draft language model (DLM) to generate draft tokens, which are then batch-verified by a large-scale target language model (TLM), thereby reducing costly TLM invocations and improving hardware parallel utilization. For memory-bounded mobile scenarios, a heterogeneous SoC consisting of an NPU and Processing-in-Memory (PIM)~\cite{attacc, maestro} is an ideal platform for implementing speculative decoding: the NPU is suitable for executing the high compute-density calculations, while PIM can efficiently handle the memory-intensive workloads.

To enhance the performance of speculative decoding in heterogeneous systems, existing research explores two primary approaches: \textbf{(i)} Operator-level synchronous partitioning~\cite{neupims,specpim} maps different operators to either the NPU or PIM and executes them in parallel synchronously at the operator granularity under a fixed draft length to balance task load and improve overall throughput; \textbf{(ii)} Task-level asynchronous scheduling~\cite{amusd} attempts to enable DLM and TLM to advance independently at their own pace on different devices, improving system parallelism and computing power utilization. However, both methods have limitations when facing adaptive drafting~\cite{specasr,svip,banditSpec,adaedl,specdec}. Synchronous partitioning assumes a fixed draft length, failing to consider the dynamic fluctuation of draft length at runtime, leading to drastic fluctuations in the load of NPU and PIM, causing mutual waiting. Asynchronous scheduling does not fully consider the matching of DLM and TLM computing characteristics with hardware computing power characteristics, and uncontrolled look-ahead drafting results in a large number of low-acceptance drafts, resulting in a waste of computational power.

\begin{figure}[t]
    \centering
    \includegraphics[width=0.47\textwidth]{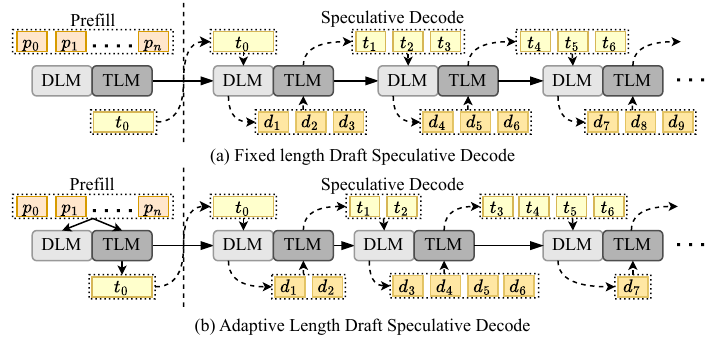}
    \vspace{-14pt}
    \caption{Speculative decoding and adaptive drafting.}
    \Description{Speculative decoding and adaptive drafting.}
    \label{fig:SpeculativeDecoding}
  \end{figure}

To address this, we propose AHASD, an asynchronous heterogeneous architecture for mobile single NPU–PIM tailored to LLM speculative decoding. AHASD achieves efficient task-level asynchronous parallelism despite computational power differences between the NPU and PIM and the category of adaptive drafting algorithms. At the architectural level, it decouples drafting and verification, enabling these tasks to be executed asynchronously on the PIM and NPU according to their computational characteristics. At the algorithm-hardware co-design level, hardware-level drafting and pre-verification controls provide learnable decision-making for switching between drafting and pre-verification. At the in-memory computing level, AHASD integrates an Attention Algorithm Unit and a Gated Task Scheduling Unit within the LPDDR5-PIM rank, enabling attention link localization and sub-microsecond task switching. The main contributions of this paper are as follows:

\textbf{(i)} We combine mobile heterogeneous systems and task-level asynchronous execution for speculative decoding based on existing adaptive drafting algorithms, mapping computing units according to the different computational characteristics of DLM and TLM, and using asynchronous queue communication and PIM computing function extensions to improve inference performance.

\textbf{(ii)} We design a hardware-oriented fusion of historical draft average entropy and leading depth online learning predictor to dynamically control adaptive drafting algorithm execution to suppress look-ahead drafting based on low-confidence drafts.

\textbf{(iii)} We model based on NPU/PIM bi-directional latency to enable the computing system to adaptively insert small-batch pre-verification, improving PIM effective compute utilization and overall speculative decoding performance.

\textbf{(iv)} We implement and evaluate AHASD end-to-end on a cycle-accurate simulator\footnote{Our simulator is open-sourced at \url{https://github.com/MAdrid1011/AHASD.git}}. Experiments with different LLMs and adaptive drafting algorithms show AHASD achieves up to 4.2$\times$ higher throughput and 5.6$\times$ greater energy efficiency than a GPU-only baseline, and 1.5$\times$ throughput and 1.24$\times$ energy efficiency improvements over the state-of-the-art GPU+PIM baseline.

%% file: sections/section2.tex
\section{Background}

\paragraph{\textbf{Speculative decoding and adaptive drafting algorithms}}


Large language models often use autoregressive decoding for output generation, relying on the previous token in each step. This makes inference process sequentially dependent, hindering throughput improvement through batch processing. Speculative decoding reduces calls to the TLM by using a smaller DLM to generate multiple candidate drafts in advance. The TLM then verifies the prefix matching degree of these candidate tokens in one forward propagation, as shown in Figure~\ref{fig:SpeculativeDecoding}(a). Early speculative decoding methods~\cite{decoding, draftVerify} used a fixed draft length. However, input context predictability varies with dialogue depth, a fixed draft length can lead to redundant drafting and verification load. Subsequent work~\cite{svip,specdec} introduced adaptive drafting algorithms to adjust the draft length at runtime based on statistical indicators, as shown in Figure~\ref{fig:SpeculativeDecoding}(b). This allows the DLM to pause drafting when drafting confidence decreases, reducing effective computing power waste.  

\begin{figure}[t]
  \centering
  \includegraphics[width=0.47\textwidth]{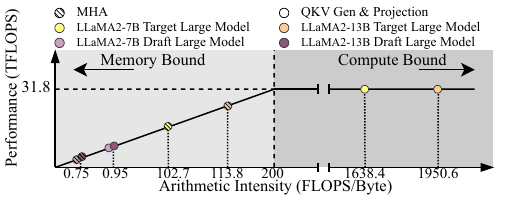}
  \vspace{-14pt}
  \caption{Roofline model of speculative decoding.}
  \Description{Roofline model of speculative decoding.}
  \label{fig:Roofline}
  \vspace{-15pt}
\end{figure}

\paragraph{\textbf{Computational characteristics of the speculative decoding}}

Speculative decoding exhibits different computational characteristics in drafting and verification: the former is memory-intensive, while the latter is compute-intensive. To verify this, we performed a roofline analysis of LLaMA2~\cite{llama2}'s inference on an NVIDIA RTX 4090 (Laptop) GPU, as shown in Figure~\ref{fig:Roofline}. The analysis indicates that: \textbf{(i)} DLM drafting mainly consists of GEMV and small-scale GEMM, with frequent parameter and cache accesses, resulting in low arithmetic intensity and memory-bound behavior; \textbf{(ii)} TLM verification can process multiple tokens in parallel, with a computational structure dominated by large-scale GEMM, resulting in high arithmetic intensity and compute-bound behavior. This difference leads to different characteristics in operator size, bandwidth requirements, and data access patterns between the two stages. Therefore, the memory-efficient and compute-efficient units can be collaboratively exploited to leverage their performance advantages.

\paragraph{\textbf{Mobile NPU–PIM heterogeneous system}}

Recent studies have examined deploying speculative decoding on heterogeneous hardware to meet its varying computational demands. In memory-constrained mobile devices, SoCs commonly use NPU-centric compute cores combined with PIM~\cite{facil}. NPUs offer high compute density, suitable for compute-intensive TLM operators like GEMM. PIM, located within the storage medium, provides high internal bandwidth, ideal for memory-intensive DLM operators. Some approaches~\cite{neupims, specpim} divide speculative decoding at the operator level, mapping different operators to distinct compute units. During each drafting and verification, DLM and TLM operators synchronize across devices based on predefined dependencies.

%% file: sections/section3.tex
\section{Motivation}

Based on the analysis above, we identify two key system-level challenges associated with running adaptive drafting speculative decoding on mobile heterogeneous systems.

\paragraph{\textbf{Challenge 1: Operator-level synchronization causes idle overhead under draft fluctuation on mobile NPU-PIM systems}}

\begin{figure}[t]
  \centering
  \includegraphics[width=0.47\textwidth]{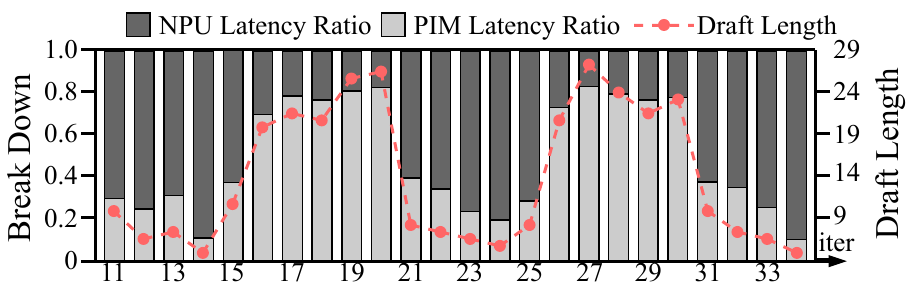}
  \vspace{-14pt}
  \caption{Overhead imbalance in operator-level scheduling.}
  \Description{Operator-level synchronization causes load imbalance under fluctuating draft lengths.}
  \label{fig:Challenge1}
  \vspace{-15pt}
\end{figure}

Current mobile NPU-PIM architectures use operator-level synchronous parallel scheduling, mapping DLM and TLM operators to the NPU or PIM to balance execution time, assuming a fixed draft length. Adaptive drafting algorithms, however, introduces algorithm-level runtime fluctuations in draft length. As shown in Figure~\ref{fig:Challenge1}, experiments on Coral-NPU~\cite{coralnpu} combined with LPDDR5-PIM~\cite{lpddr5}, replicating SpecPIM~\cite{specpim} task mapping with LLaMA2-1.3B (DLM) and LLaMA2-7B (TLM) using the AdaEDL~\cite{adaedl} algorithm, show that as draft length causes PIM latency to fluctuate significantly (12.3\% to 84.2\% of total inference latency) due to surging computational load, while the well-provisioned NPU remains idle, causing synchronization overhead. Conversely, shorter drafts make the NPU the bottleneck, causing both devices to frequently idle waiting for each other's results.

\paragraph{\textbf{Opportunity 1: Performance optimization potential of task-level NPU-PIM asynchronous parallelism}}

To address algorithm-level workload imbalances from draft fluctuations, scheduling must overcome operator-level synchronization constraints. Task-level asynchronous scheduling relaxes cross-device dependencies, allowing drafting and verification tasks to proceed independently and decoupling PIM-side drafting from NPU-side verification. Research indicates that DLM can generate candidate tokens without verification feedback~\cite{pearl, amusd}, supporting this asynchronous method. While operators within each task execute sequentially, overall system parallelism improves, reducing synchronization overhead.

\paragraph{\textbf{Challenge 2: Acceptance rate degradation of look-ahead drafting leads to effective computing power waste}}

\begin{figure}[t]
  \centering
  \includegraphics[width=0.47\textwidth]{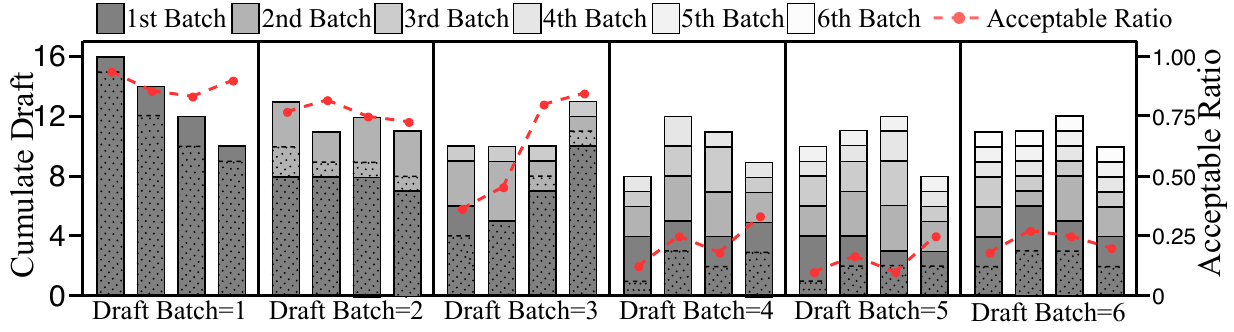}
  \vspace{-13pt}
  \caption{The acceptable ratio of look-ahead drafting.}
  \Description{The Low Acceptable Ratio of Look-ahead Drafting}
  \label{fig:Challenge2}
  \vspace{-15pt}
\end{figure}

Decoupling the DLM and TLM in task-level asynchronous scheduling enhances system parallelism but causes dynamic load drift between the NPU and PIM. In strongly context-constrained stages, the adaptive algorithm produces longer drafts, extending the DLM’s single-batch computation time. Meanwhile, the TLM continuously verifies unverified tokens in parallel, minimizing overall performance impact. In weakly context-constrained stages, shorter drafts reduce the DLM’s computation time, while the TLM load remains stable. To maintain utilization, the DLM continues drafting based on unverified tokens; however, the acceptance rate of drafts generated under weak context constraints is naturally low, causing multiple draft batches to be rejected by the TLM. Empirical measurements show that in mobile scenarios, when LLaMA2-1.3B (DLM) and LLaMA2-7B (TLM) are deployed on the Coral-NPU + LPDDR5-PIM platform, as shown in Figure~\ref{fig:Challenge2}, a decrease in the single-batch quantity of adaptive drafts leads to a surge in the number of unverified draft batches between adjacent verification cycles, resulting in an insufficient number of accepted tokens. Further drafting by the DLM based on low-confidence drafts causes significant waste of effective computing power on the PIM side.

\paragraph{\textbf{Opportunity 2: Potential gains from PIM-side small-batch pre-verification}}

As shown in Figure~\ref{fig:Challenge2}, short adaptive draft lengths enable small-batch pre-verification of the earliest unverified tokens in look-ahead drafts, more accurately reflecting the overall acceptance trend. This low-intensity pre-verification can be efficiently executed on PIM using GEMV. Consequently, TLM can be triggered by the PIM to perform small-batch pre-verification on unverified tokens, preventing invalid look-ahead drafting around low-confidence drafts. However, if pre-verification delays exceed the verification time on the NPU, draft exhaustion may occur, causing the NPU to idle after completing current verification. To address this, the system must model execution times across computing units, perform online runtime estimation of DLM and TLM on PIM and NPU, and determine the optimal timing for switching between drafting and pre-verification, as well as the pre-verification length on the PIM. 

%% file: sections/section4.tex
\section{AHASD Design}


To address these challenges, we propose AHASD, a mobile NPU-PIM heterogeneous architecture for LLM speculative decoding, featuring three key mechanisms. First, to tackle the inefficiency of operator-level synchronization (Challenge 1), AHASD allocates computing units based on the distinct computational characteristics of DLM and TLM. As draft length varies dynamically during adaptive decoding, each device operates asynchronously, thereby avoiding mutual waiting and idling that typically occur with conventional operator-level synchronization. Second, to address the fundamental limitations caused by excessive low-confidence in look-ahead drafting (Challenge 2), AHASD introduces Entropy-History-Aware Drafting Control. This method, based on the model’s inference state, combines historical prediction entropy with the number of leading draft batches relative to verification, enabling hardware-level online learning, effectively suppressing further look-ahead drafting based on low-acceptance drafts. Furthermore, also aiming for the Challenge 2, AHASD employs Time-Aware Pre-Verification Control. This approach, based on the execution state, integrates runtime latency modeling of both the NPU and PIM, enabling the insertion of small-batch pre-verification tasks on the PIM without causing NPU idling, thereby improving computational utilization. 

\begin{figure}[t]
  \centering
  \includegraphics[width=0.45\textwidth]{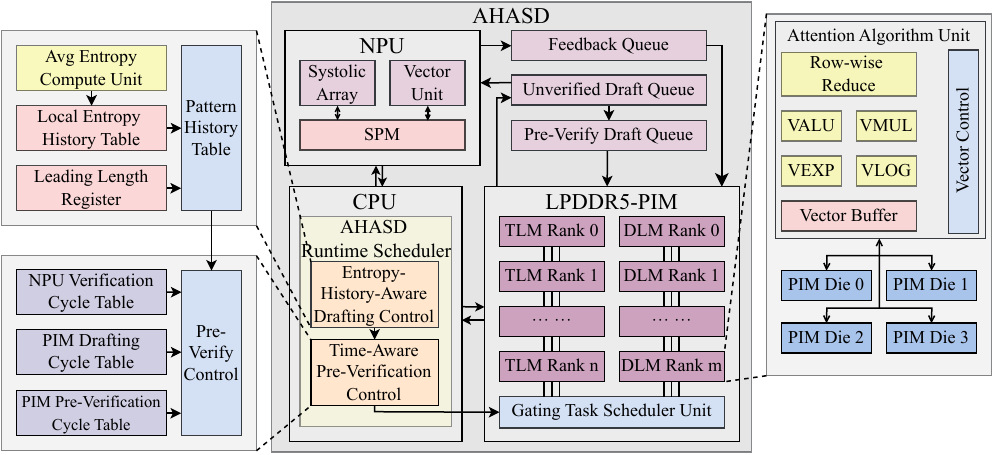}
  \vspace{-10pt}
  \caption{The architecture design of AHASD.}
  \Description{The system architecture of AHASD.}
  \label{fig:Design}
  \vspace{-15pt}
\end{figure}

Figure~\ref{fig:Design} shows the AHASD system architecture, integrating a mobile NPU with LPDDR5-PIM memory, while the CPU handles task scheduling and control. The CPU and NPU communicate with the LPDDR5 module via a high-speed data bus. For algorithm data flow, asynchronous queues enable decoupled cross-device data transfer and task-level parallelism, allowing asynchronous execution of DLM and TLM. The NPU, resembling a mobile SoC, comprises a systolic array, vector unit, and on-chip SPM. The LPDDR5-PIM module features multiple memory channels supporting in-memory computing to maximize limited edge device bandwidth.


\subsection{AHASD Task-Level Asynchronous Heterogeneous Execution Framework}

To address workload imbalances from adaptive drafting, AHASD assigns memory-intensive drafting to PIM and compute-intensive verification to NPU, enabling asynchronous task collaboration. It uses three cross-device asynchronous queues: \textbf{(i)} an unverified draft queue holding token batches from PIM awaiting NPU verification; \textbf{(ii)} a feedback queue storing TLM verification results to guide PIM in confirming or rolling back drafts; and \textbf{(iii)} a pre-verification queue managed by the CPU scheduler to mark drafts needing pre-verification in PIM. These queues facilitate data exchange between DLM and TLM while ensuring result consistency. 

Regarding in-memory computing support, AHASD integrates an Attention Algorithm Unit (AAU) within each LPDDR5-PIM rank. The AAU executes nonlinear operators as well as reduction operations, directly on the in-memory data path. This in-situ processing eliminates the need to transfer intermediate activations to the NPU, effectively reducing cross-chip communication overhead.

Upon detecting pre-verification, the Gated Task Scheduling Unit enables computation units in the ranks with TLM parameters while disabling those in the DLM ranks. By leveraging rank-level gating, it achieves sub-microsecond switching, allowing the PIM to execute pre-verification without wasting computation time.

\subsection{Entropy-History-Aware Drafting Control}

\begin{figure}[t]
  \centering
  \includegraphics[width=0.40\textwidth]{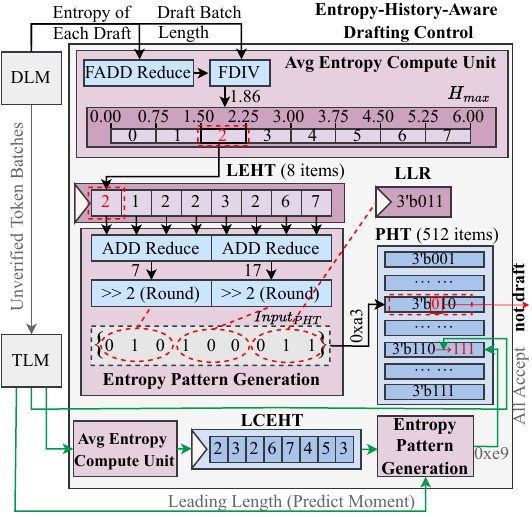}
  \vspace{-14pt}
  \caption{Entropy-History-Aware Drafting Control module.}
  \Description{Entropy-History-Aware Drafting Control Module.}
  \label{fig:edc}
  \vspace{-15pt}
\end{figure}

Based on the current DLM's inference confidence, the Entropy-History-Aware Drafting Control (EDC) module, proposed by our AHASD, determines whether to continue look-ahead drafting by considering the model's inference state, thereby enhancing the effective computational throughput of the PIM.

Figure~\ref{fig:edc} illustrates the EDC module along with a simple example. After each batch of drafting is completed, AHASD's PIM calculates the average softmax entropy value $\overline{H}$ of the batch of tokens and maps it to one of eight equally spaced discrete intervals within the range $[0, H_{\max}]$, where $H_{\max}$ is statically preset by the statistical maximum softmax entropy observed during large model inference. The CPU writes the mapped bucket number to the Local Entropy History Table (LEHT) inside the EDC and increments the 3-bit Leading Length Register (LLR), which records the number of unverified draft batches currently leading the verification. To capture dynamic changes in drafting, the EDC divides the LEHT into two groups ($H_{0-3}, H_{4-7}$), calculates the average entropy within each group, and concatenates these averages to form the historical entropy feature $\{\overline{H_{4-7}}, \overline{H_{0-3}}\}$. Subsequently, the LLR is appended as the low bit to form the 9-bit input index for the Pattern History Table (PHT):
\[
Input_{PHT} = \{\overline{H_{4-7}},\ \overline{H_{0-3}},\ LLR\}
\]

By jointly indexing these factors, the PHT can capture the evolving correlation between entropy fluctuations and draft acceptance. When the highest bit of the pattern output by the PHT is 1, it indicates that the batch of drafts is judged to have a high acceptance potential and is worth continuing to generate in advance.

After the NPU completes verification, the CPU immediately updates the state of the EDC by reducing the LLR count and using the average entropy calculation unit for the accepted batches to write the average entropy to the Local Commit Entropy History Table (LCEHT). If a draft batch is rejected, the CPU rolls back the LCEHT content to the LEHT. The update of the PHT depends solely on the LEHT and the current verification result: if the draft is fully accepted, the corresponding item counter is incremented; otherwise, it is decremented. This process gradually learns whether the entropy pattern of the current draft will be verified by the TLM.

\subsection{Time-Aware Pre-Verification Control}

\begin{figure}[t]
  \centering
  \includegraphics[width=0.42\textwidth]{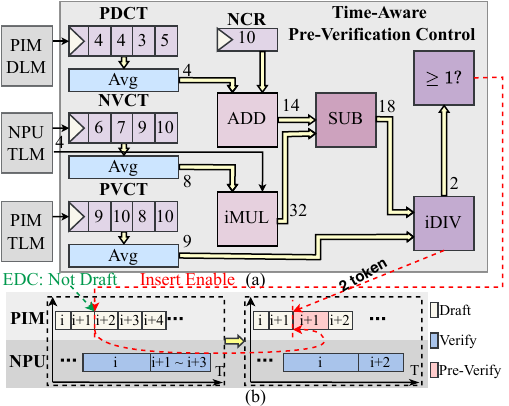}
  \vspace{-14pt}
  \caption{Time-Aware Pre-Verification Control module.}
  \Description{The Time-Aware Verify Control Module.}
  \label{fig:tvc}
  \vspace{-15pt}
\end{figure}

After EDC recommends whether to continue speculative drafting from a model inference perspective, the system must decide if small-batch pre-verification should begin on the PIM side without adding synchronization overhead between the NPU and PIM. To address this, AHASD uses the Time-Aware Pre-Verification Control (TVC) module, which performs runtime two-sided latency modeling, as shown in Figure~\ref{fig:tvc}(a). For the NPU, execution latency is mainly limited by parameter transfer overhead. The KV cache transfer caused by generated tokens impacts execution time. Thus, TVC predicts the execution cycles for the NPU's current (i-th) task as:
\[
C_{NPU_i}=\frac14\sum_{j=0}^3(\frac {C_{NPU}}{L_{KV}})_{j}\times L_{KV_i}
\]
where, $(\frac{C_{NPU}}{L_{KV}})_j$ is the ratio of the equivalent number of NPU execution cycles—converted to the PIM side according to the frequency ratio of PIM and NPU—recorded by the NPU-side Verification Cycle Table (NVCT) to the KV cache length involved in the inference, and $L_{KV_i}$ is the KV cache length used in this inference. Since PIM is computation-limited, task delay is nearly linearly related to batch processing length. Our DLM time prediction model mirrors that used for the NPU. The delays for drafting and pre-verification are:
\[
C_{PIM-Draft_i} =\frac14\sum_{j=0}^3(\frac {C_{PIM-DLM}}{L_{Draft}})_j\times L_{Draft_{i}}
\]\[
C_{PIM-Verify_i} =\frac14\sum_{j=0}^3(\frac {C_{PIM-TLM}}{L_{Draft}})_j\times L_{Draft_{i}}
\]
where, $(\frac{C_{PIM-DLM}}{L_{Draft}})_j$ and $(\frac{C_{PIM-TLM}}{L_{Draft}})_j$ are the ratios of the number of execution cycles for DLM and TLM inference, respectively, as recorded by the PIM-side Drafting Cycle Table (PDCT) and the Pre-Verification Cycle Table (PVCT), to the length of the inference draft. Here $L_{Draft_i}$ is the length of the draft processed in this inference. Additionally, to ensure the stability of early predictions, TVC presets the average execution cycle of a single token—obtained from offline profiling for NVCT, PDCT, and PVCT—at the beginning of the task.

When the EDC deems drafting unbeneficial, the TVC adopts a conservative approach to decide on inserting a pre-verification on the PIM side. Specifically, before completing the current NPU verification, the PIM must \textbf{(i)} complete the pre-verification of several drafts and \textbf{(ii)} generate at least one new draft if the pre-verification result is potentially unacceptable. This ensures the NPU has new drafts available for the next verification, instead of idling. Thus, the remaining cycles available for PIM pre-verification is:
\[
C_{PIM-Left} = C_{NPU_i}-(C_{now} + C_{PIM-Draft_1})
\]
where $C_{now}$ is the equivalent NPU verification cycles recorded by the NPU Current Execution Cycle Register (NCR). Combined with the PIM-side pre-verification cycle evaluation, this yields the draft length eligible for pre-verification; TVC inserts pre-verification only when this length $\geq 1$, otherwise continuing draft generation. The modified calculation flow is shown in Figure~\ref{fig:tvc}(b).

%% file: sections/section5.tex
\section{Evaluation}

\subsection{Methodology}

\begin{table}[t]
  \centering
  \caption{Benchmark Model Configurations}
  \vspace{-10pt}
  \resizebox{0.47\textwidth}{!}{
  \begin{tabular}{lcccc}
  \toprule
  \textbf{Scale} & \multicolumn{2}{c}{\textbf{Draft Model}} & \multicolumn{2}{c}{\textbf{Target Model}} \\ 
  \cmidrule(r){2-3} \cmidrule(r){4-5}
  & \textbf{Name} & \textbf{Hidden Size} & \textbf{Name} & \textbf{Hidden Size} \\
  \midrule
  Small  & OPT-1.3B      & 2048  & OPT-6.7B      & 4096  \\
  Medium & LLaMA2-7B     & 4096  & LLaMA2-13B    & 5120  \\
  Large  & PaLM-Like-8B       & 4096  & PaLM-Like-30B      & 8192  \\ 

  \bottomrule

  \end{tabular}
  }
  \label{tab:benchmark}
  \vspace{-15pt}
\end{table}

\paragraph{\textbf{Baseline}}

We compare AHASD with two baselines: \textbf{(i) GPU-Only}, where drafting and verification are performed alternately and sequentially on the GPU; and \textbf{(ii) SpecPIM}~\cite{specpim}, a speculative decoding operator-level parallel acceleration scheme based on a GPU+PIM hybrid. We reconstruct the PIM component using the simulation methodology described in the SpecPIM paper, based on the Samsung HBM-PIM design~\cite{hbmpim} and SK-Hynix's GDDR-PIM design~\cite{gddr6}, and employ two GPUs to build its Host. The GPUs used in the baselines are NVIDIA GeForce RTX 4090 (Laptop) @1.335GHz.

\paragraph{\textbf{Benchmark}}
As shown in Table~\ref{tab:benchmark}, we evaluate AHASD's performance using three benchmark tests with different model configurations: OPT~\cite{opt}, LLaMA2~\cite{llama2}, and PaLM~\cite{palm} (PaLM-like surrogate models based on the PaLM architecture and published parameter scales). All models are quantized to INT8, with varying rank sizes stored in LPDDR5 according to model size. We apply four adaptive drafting algorithms—SpecDec++~\cite{specdec}, SVIP~\cite{svip}, AdaEDL~\cite{adaedl}, and BanditSpec~\cite{banditSpec}—to perform adaptive drafting for each model. For each benchmark, the generation length is set to 1024 and the batch size to 1, simulating the small batch sizes typical of mobile terminals, with inference performed on the Alpaca~\cite{alpaca} dataset. 

\paragraph{\textbf{Experiment platform}}

Using two clock-accurate open-source simulators, ONNXim~\cite{onnxim} and SAITPublic-PIMSimulator~\cite{pimSimulator}, we simulate mobile NPUs and PIM, respectively, with performance metrics shown in Table~\ref{tab:platform}. We modify ONNXim's memory interface and added code for three asynchronous queues to enable communication between the simulators. Furthermore, we integrate the EDC and TVC modules into the Xiangshan~\cite{xiangshan} open-source CPU-based SoC to implement dynamic scheduling of AHASD's tasks.

  \begin{figure*}[t]
    \centering
    \includegraphics[width=0.9\textwidth]{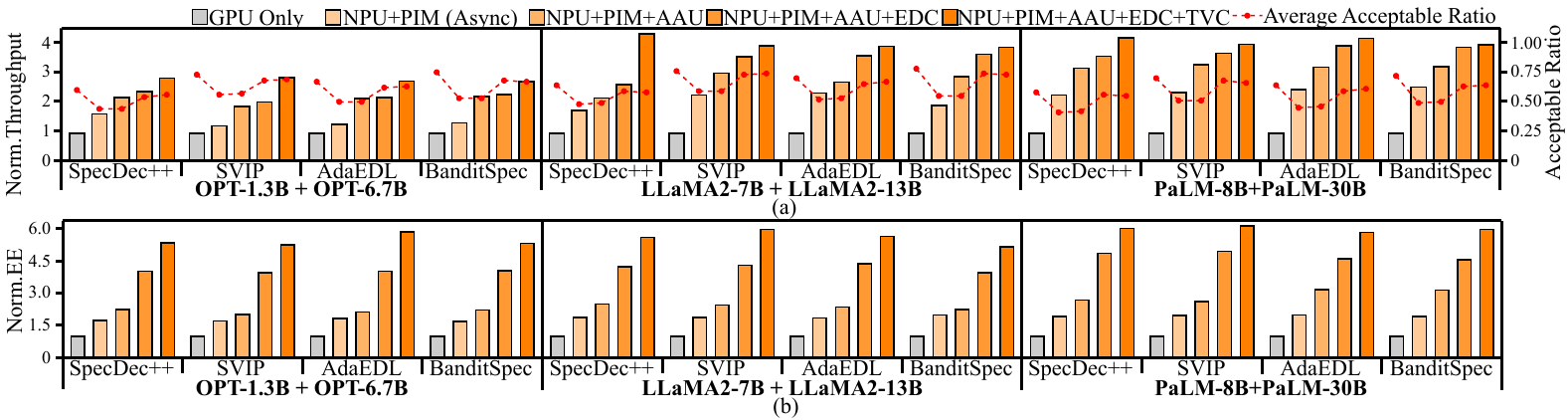}
    \vspace{-14pt}
    \caption{Results of ablation experiments.}
    \Description{Results of ablation experiments.}
    \label{fig:ablation}
    \vspace{-15pt}
  \end{figure*}

\subsection{Ablation Experiments}

\begin{table}[t]
    \centering
    \caption{Hardware Configuration of Experiment Platform}
    \vspace{-10pt}
    \resizebox{0.47\textwidth}{!}{
    \renewcommand{\arraystretch}{1.25}
    \begin{tabular}{|c|c|c|c|} 
    \hline
    \multicolumn{2}{|c|}{\textbf{Mobile NPU Configuration}} & \multicolumn{2}{c|}{\textbf{PIM (LPDDR5) Configuration}} \\ \hline
    
    Matrix compute unit & 16 TOPS (INT8) & Number of PIM units & 16 \\ \hline
    Vector compute unit & 8.2 TOPS (INT8) & PIM performance  & 102.4 GOPS (INT8) \\ \hline
    Number of compute chips & 2 & Capacity & 4 GB/rank $\times$16 \\ \hline
    Operating frequency & 1 GHz & On-chip bandwith & 256 GB/s  \\ \hline
    Scratchpad Capacity& 8MB & Off-chip bandwidth & 51.2 GB/s\\ \hline
    
    \multicolumn{4}{|c|}{\textbf{PIM (LPDDR5) Timing Parameters}} \\ \hline
    \multicolumn{4}{|c|}{
    $t_{RP}=32,\;
    t_{RCD}=32,\;
    t_{RAS}=64,\;
    t_{RRD_L}=8,\;
    t_{WR}=24,$
    } \\
    
    \multicolumn{4}{|c|}{
    $t_{CCD_S}=4,\;
    t_{CCD_L}=6,\;
    t_{REFI}=6240,\;
    t_{FAW}=64,\;
    t_{RFC}=560$
    } \\ \hline
    
    \end{tabular}
    } 
    \label{tab:platform}
    \vspace{-15pt}
    \end{table}



\paragraph{\textbf{Throughput and average draft acceptance rate}}


As shown in Figure~\ref{fig:ablation}(a), NPU+PIM task-level asynchronous scheduling reduces the average acceptance rate by 25.1\% due to drafting based on unverified tokens, yet raises throughput to 2.2$\times$ on average, because the DLM can generate drafts asynchronously and independently of the TLM. Adding AAU completes non-linear and reduction operations on-chip, cutting communication overhead and lifting throughput to 2.7$\times$. Further introducing EDC suppresses look-ahead drafting on low-confidence tokens, recovering 24.6\% in acceptance rate and reaching 3.4$\times$—though gains vary by model and algorithm (e.g., LLaMA2+AdaEDL achieves 33\% acceleration over NPU+PIM+AAU, while OPT improves by less than 10\%). Finally, TVC dynamically controls pre-verification insertion based on execution time, reducing draft exhaustion and NPU idling; despite slightly perturbing EDC, it brings final throughput to 3.8$\times$ on average.


\paragraph{\textbf{Energy efficiency (EE)}}

As shown in Figure~\ref{fig:ablation}(b), the task-level asynchronous NPU+PIM architecture significantly reduces synchronization overhead caused by mutual waiting by decoupling generation and verification. However, this approach introduces some overhead due to reduced draft acceptance rates, resulting in an average energy efficiency improvement of 1.9$\times$. Furthermore, integrating the AAU within the PIM to reduce on-chip transfer of intermediate results from attention computation—although it introduces additional energy consumption overhead from the AAU—consistently improves energy efficiency to an average of 2.6$\times$. After introducing EDC, the average energy efficiency increases to 4.5$\times$, attributed to EDC's effective suppression of invalid computational waste caused by low-confidence drafts, which reduces rollback overhead in the PIM. Finally, incorporating the TVC module adds only the minimal necessary pre-verification within the NPU execution window, further reducing NPU idle time caused by the lack of drafts to verify, resulting in an overall average energy efficiency of 5.2$\times$.

\subsection{Comparison with State-of-the-Art Methods}


\paragraph{\textbf{Throughput}}

\begin{figure}[t]
  \centering
  \includegraphics[width=0.47\textwidth]{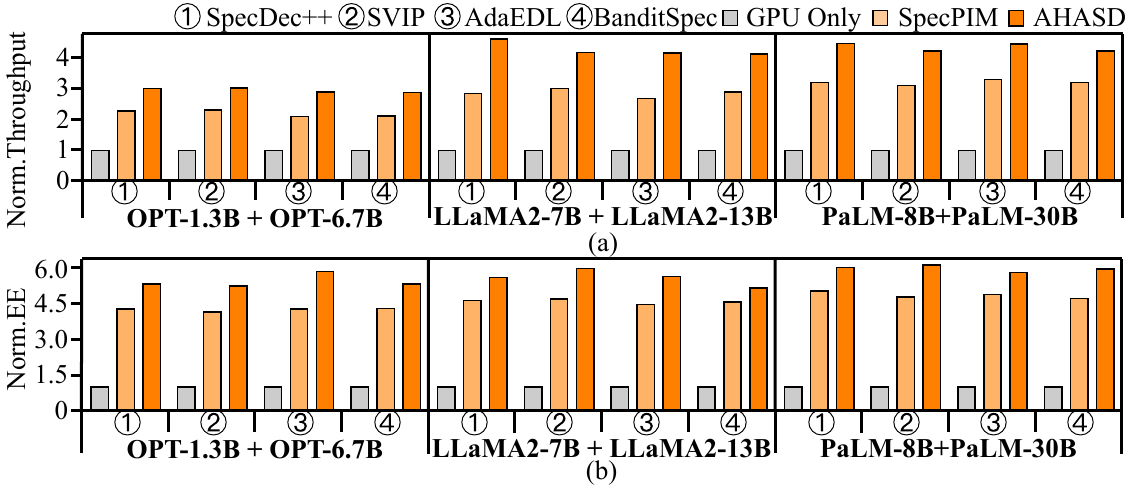}
  \vspace{-14pt}
  \caption{Results of comparison with state-of-the-art.}
  \Description{Results of comparison with state-of-the-art.}
  \label{fig:SOTA}
  \vspace{-15pt}
\end{figure}

Figure~\ref{fig:SOTA}(a) compares the throughput of our architecture with that of a GPU and SpecPIM. AHASD achieves up to 4.2$\times$ higher throughput than the GPU by employing adaptive drafting reasoning, offloading low-compute-intensity operators in DLM to the PIM, and reducing data transfer time. It also outperforms SpecPIM (GPU+PIM) by up to 1.5$\times$, mainly due to task-level asynchronous scheduling and look-ahead drafting control. While SpecPIM balances computation time between the PIM and GPU through design space exploration, its operator-level parallelism introduces overhead imbalances when handling adaptive drafting tasks with algorithm-determined draft lengths. AHASD's task-level asynchronous scheduling effectively mitigates this issue. Additionally, as shown in Figure~\ref{fig:ablation}(a), AHASD's look-ahead drafting and pre-verification control prevents low draft acceptance rates.

\paragraph{\textbf{Energy efficiency}} 

Figure~\ref{fig:SOTA}(b) shows the energy efficiency of AHASD compared to the GPU and SpecPIM. AHASD achieves up to a 5.6$\times$ increase in energy efficiency relative to the GPU, primarily due to reduced data transfer and communication overhead. Compared to architectures like SpecPIM, which fully optimize energy efficiency through design space exploration, AHASD attains an average energy efficiency improvement of approximately 1.24$\times$. This enhancement is mainly attributed to the reduced synchronous idling of the NPU and PIM enabled by asynchronous scheduling, as well as decreased communication overhead during non-linear operator calculations facilitated by the AAU.

\subsection{Overhead Analysis}

\paragraph{\textbf{Area overhead}}

The primary overhead of AHASD resides in the EDC, TVC, AAU, and NPU-PIM asynchronous queues. To evaluate their hardware cost, we used CACTI~\cite{cacti} for SRAM area estimation and Yosys combined with OpenROAD for logic synthesis and area estimation. All area estimates are based on the 28 nm process node. Table~\ref{tab:area} presents the area estimation results for these four modules. It can be observed that the EDC, TVC, and asynchronous queues mainly consist of small-scale registers with minimal control logic, totaling less than 0.09 mm$^2$, which accounts for under 0.2\% of the LPDDR5-PIM die. The AAU synthesis area is 1.25 mm$^2$, representing no more than 2.5\% of the LPDDR5-PIM die. Overall, the additional hardware area overhead of AHASD is 2.68\% of the DRAM, which is much lower than that of DRAM peripherals and PIM logic.

\paragraph{\textbf{Power overhead}}

To evaluate AHASD's power performance, we measured memory-side background power (including clock, I/O, row buffer maintenance, etc.) and dynamic power using the LPDDR5 power model. The AAU and small-batch pre-verification operators were converted into equivalent read-modify-write energy. On the NPU side, we included DMA transfer, SPM access, and computing core dynamic power during TLM verification. The results are presented in Table~\ref{tab:power}. Compared to the baseline (NPU + LPDDR5), AHASD introduces additional memory access operations due to asynchronous queues and the AAU, resulting in an average increase of approximately 2.00$\times$ in LPDDR5 memory-side power consumption. However, benefiting from task-level asynchrony and EDC/TVC's invalid drafting suppression, the end-to-end throughput increases by an average of 3.10$\times$, leading to an overall energy consumption reduction of approximately 50\%. Compared to heterogeneous architectures with added PIM, AHASD's LPDDR5 power consumption is similar, but look-ahead drafting control and pre-verification control reduce idling and improve throughput by 1.41$\times$, yielding a 25\% increase in energy efficiency.


\begin{table}[t]
    \centering
    \caption{Area Overhead Summary}
    \vspace{-10pt}
    \resizebox{0.47\textwidth}{!}{
    \begin{tabular}{lcc}
    \hline
    \textbf{Module} & \textbf{Area (mm$^2$)} & \textbf{Percent of LPDDR5-PIM Die} \\
    \hline
    EDC & 0.05 & $ 0.10\%$ \\
    TVC                  & 0.03 & $ 0.06\%$ \\
    Async Queue             & 0.01 & $ 0.02\%$ \\
    AAU      & 1.25  & $ 2.50\%$ \\
    \hline
    \end{tabular}
    \label{tab:area}
    }
    \vspace{-10pt}
  \end{table}

\begin{table}[t]
  \centering
  \caption{Power \& Efficiency Comparison}
  \vspace{-10pt}
  \resizebox{0.47\textwidth}{!}{
  \begin{tabular}{lcccc}
  \hline
  \multirow{2}{*}{\textbf{Configuration}} & \multirow{2}{*}{\textbf{LPDDR5}} & \multirow{2}{*}{\textbf{NPU}} & \multirow{2}{*}{\textbf{Throughput}} & \textbf{Energy} \\
  & & & & \textbf{per token} \\
  \hline
  NPU + LPDDR5 (base)  & 1.00$\times$          & 1.00$\times$          & 1.00$\times$        & 1.00$\times$        \\
  NPU + LPDDR5-PIM  & 1.90$\times$    & 1.10$\times$    & 2.20$\times$    & 0.64$\times$  \\
  AHASD        & 2.00$\times$    & 1.15$\times$    & 3.10$\times$    & 0.48$\times$  \\
  \hline
  \end{tabular}
  }
  \label{tab:power}
  \vspace{-15pt}
  \end{table}

%% file: sections/section6.tex
\section{Conclusion}


This paper presents AHASD, an asynchronous heterogeneous speculative decoding architecture designed to address load imbalance and computational inefficiency in adaptive drafting speculative decoding on mobile NPU–PIM architectures. AHASD separates the DLM and TLM components using a task-level asynchronous framework and incorporates Entropy-History-Aware Drafting Control combined with Time-Aware Pre-Verification Control to enhance inference efficiency. Experimental results demonstrate that AHASD improves throughput by up to 4.2$\times$ and energy efficiency by up to 5.6$\times$ compared to GPUs. It also maintains a throughput advantage of up to 1.5$\times$ and an energy efficiency advantage of up to 1.24$\times$ over SpecPIM, with hardware overhead below 3\% of the DRAM die area, indicating strong potential for lightweight mobile deployment. 